# Trustable Mobile Crowd Sourcing for Acquiring Information from a Flooded Smart Area

Sajedeh Abbasi
PerLab, Faculty of Electrical and Computer Engineering, University of Birjand, Birjand, Iran
sajedeh_abbasi@birjand.ac.ir

Hamed Vahdat-Nejad[*]
PerLab, Faculty of Electrical and Computer Engineering, University of Birjand, Birjand, Iran
vahdatnejad@birjand.ac.ir

Hamideh Hajiabadi
Department of Computer Engineering, Birjand University of Technology, Iran
hajiabadi@birjandut.ac.ir

*Abstract*— **Flood is a natural phenomenon that causes severe environmental damage and destruction in smart cities. After a flood, topographic, geological, and living conditions change. As a result, the previous information regarding the environment is no more valid. Rescue and relief organizations that intend to help the affected people need to obtain new and accurate information about the conditions of the flooded environment. Acquiring this required information in the shortest time is a challenge for realizing smart cities. Due to the advances in the Internet of Things technology and the prevalence of smartphones with several sensors and functionalities, it is possible to obtain the required information by leveraging the Crowdsourcing model. In this paper, the information required from a flooded area is classified into four categories: victim, Facility and Livelihood, medical, and transfer. Next, a crowdsourcing scheme for acquiring information is proposed, including malicious user detection to ensure the accuracy of information received. Finally, simulation results indicate that the proposed scheme correctly detects malicious users and ensures the quality of obtained information.**

## I. INTRODUCTION

Flood is the most common natural disaster, accounting for more than half of the casualties and one-third of the economic damages caused by natural disasters [1]. Disaster management includes four main steps: prevention, preparedness, response, and recovery [3]. Following a disaster (in the response phase), various agencies and relief forces volunteer and work together to help people in the affected regions [2]. In order to help appropriately, these teams need accurate information in the shortest time [4].

With the advent of the Internet of Things (IoT) and smart cities, several daily functionalities have become intelligent. In particular, the fantastic capabilities of smartphones, such as high-quality cameras, internet connection, high-speed processors as well as the presence of various sensors such as GPS, have made them a practical device for information acquisition. Due to the prevalence of mobile phones, the Crowdsourcing paradigm has been proposed in smart cities to obtain the information needed by the relief forces [5, 6]. As the people available in the disaster area have accurate information about the situation in the region, they can get involved in the information acquisition process via mobile crowdsourcing [7].

In this regard, the MAppERS application [8] provides a platform in which users enter the water level and upload photos taken from the height of the water surface. They can also mark their requirements in the list. Similarly, in the Amrita Kripa system [9], users can request food, warm clothes, health supplies, and medical care and also submit their service requests such as moving appliances or cleaning the house.

One of the main challenges of Crowdsourcing systems is detecting malicious users who corrupt the whole data [10]. Specifically, in disaster management, incorrect information misleads rescue workers and can lead to severe damages. In this regard, this paper extends state of the art by proposing a malicious user detection algorithm for the proposed crowdsourcing-based information acquisition system. We simulate five types of users and validate the malicious user detection algorithm.

The structure of the rest of the paper is as follows. In Section 2, the analysis of the information needs of the data collection system is performed. Section 3 describes the proposed malicious user detection algorithm. Section 4 evaluates the proposed algorithm, and Section 5 concludes the paper.

## II. INFORMATION NEEDS ASSESSMENT

After a disaster, various relief teams enter the area to help the injured people. To meet the information needs of the relief forces, we have reviewed the needs of the flood-stricken people [8, 9, 11] as well as the Red Cross and Red Crescent websites. Finally, to meet these needs accurately, we have interviewed experts from the Iranian Red Crescent Society. Then, the information needed for the relief forces is divided into four categories, including "Victim", "Facility and Livelihood", "medical", and "transfer".

- Victim: As a result of a severe flood, some people would lose their lives, some would be injured, and many would be missing. Rescue workers need information regarding the dead, injured, and missing people to help the victims. They also need to know the remained population in the region to send supplies.

- Facility and livelihood: Water generally floods homes in flooded regions and causes many problems. Access to food, baby formula, drinking water, tents, blankets, warm clothing, diapers, and toiletries are among the most important needs of flood victims, which relief teams should provide.

- Medical: The injured should be provided with medical facilities and doctors. Some people affected by the flood might have certain illnesses and take certain medications that they do not have access to. Therefore, the rescue forces must have information about the medicines needed by the people and the injured.

- Transfer: One of the important needs of flooded people is to transfer to a safe place. Typically, houses in a flooded region might become uninhabitable. Relief teams need information on uninhabitable houses to prepare a safe place with adequate facilities for the evacuation. In some cases, a house is uninhabitable. Still, people refuse to transfer to a safe place due to the presence of an elderly or disabled person or because they have livestock or pets that they are unable to carry to a safe place. As a result, rescue workers need information about the elderly, the disabled, livestock, or pets. Rescuers also need information on the flooding of roads and streets to determine how to provide relief.

After acquiring and categorizing the needs, a questionnaire including 15 multiple-choice and two descriptive questions is designed to be provided to flooded users under one application. The application has an optional file upload capability (audio, photo, and video) for each question so that users can send audio, photo, or video to augment the information of each question. It is also possible for users to skip a question. When users log in to the system, they also enter their identity information, level of education, passed courses related to relief, and user experience in past information needs collection operations. After answering the questions, the answers are uploaded to the server. The server implements the proposed malicious user detection algorithm, and then the needs of the people in each area will reach the corresponding organizations and relief forces.

### III. MALICIOUS USER DETECTION ALGORITHM

In some cases, the Crowdsourcing system is exposed to inaccurate information, making the whole system un-trustable. To resolve this issue, Crowdsourcing systems need to detect intentional or non-intentional malicious users. To this end, a malicious user detection algorithm is proposed, which consists of two parts, including outlier data detection and malicious user detection. In this algorithm, the statistical approach is used to detect the outlier data, and then malicious users are identified and removed from the system.

The disaster area is partitioned into R regions to identify users' needs accurately, and the algorithm is run separately for each region. Each user's answer to each question is converted to a number. For example, for two-choice questions, the first option is assigned the number 1, and the second option is assigned the number 2. For the three-choice questions, the first option is assigned the number 1, the second option is assigned the number 2, and the third option is assigned the number 3, and the same is done for other types of questions. Afterward, for each period (i.e., one hour), the average and standard deviation of the values of each question are calculated per region, as follows:

$$\bar{x} = \left(\frac{1}{N}\right)\sum_{i=1}^{N} x_i \quad (1)$$

$$\delta = \sqrt{\left(\frac{1}{N}\right)\sum_{i=1}^{N}(x_i - \bar{x})^2} \quad (2)$$

Where N is the number of data elements entered for that question, $x_i$ indicates received values, $\bar{x}$ is the mean value, and $\delta$ is the standard deviation [12]. After calculating the mean and standard deviation of the data received in the last period for a specific region, the interval $[\bar{x}-2\delta, \bar{x}+2\delta]$ is considered as the valid interval for the mentioned period and region. This means that if the user's data is placed within this interval, the user's participation in that question is considered valid (useful). Otherwise, it is considered outlier participation. The number of outlier data for each user is then counted.

To avoid an adversary dominating the mean value by providing many false participations, the mean and deviation are calculated based on per-user participation, which means that only one value is considered for each user. To this end, if a user participates more than once during the last period, the average data sent by them for each question is calculated and is regarded as the value entered by that user in the last period. Then, relations (1) and (2) will be applied; These relationships must be calculated for each question.

After determining the number of outlier data provided by a user, the type of user participation (Malicious or non-Malicious) is determined by Equation (3).

$$A_i = \frac{\text{Number of outlier Data Elements for user } i}{\text{Total Sent Data Elements by user } i} \quad (3)$$

participation type: $\begin{cases} A_i <= 0.5 & \text{non-Malicious User} \\ A_i > 0.5 & \text{Malicious User} \end{cases}$

The variable $A_i$ shows the rate of outlier data provided by the $i_{th}$ user. According to Equation (3), a malicious user is a user whose at least 50% of the submitted data is outlier data. If a malicious user is detected, all information sent by them will be completely removed from the system, and the user will not be considered in future participations. By removing the malicious users and the data they have sent, the aggregated information that reaches the rescue forces is more reliable.

The malicious user detection algorithm is executed periodically in one-hour intervals. To reflect the minimum number of statistical instances for calculating mean and standard deviation, the malicious user detection algorithm is executed if at least five users have participated in the system in the last period. Otherwise, the outlier data and the malicious users cannot be correctly identified in this round.

## IV. EXPERIMENT

A simulator has been designed using MATLAB software to evaluate the proposed method. In this regard, five types of users are simulated.

User 1 (random) provides the answers completely randomly.

User 2 (pattern) selects the answers with a specific pattern, for example, for question 1 option 1, for question 2 option 2, for question 3 option 3, etc. In fact, the options are selected in rotation.

User 3 (accurate) provides the answer to all the questions correctly.

User 4 (Normal distribution with low variance) answers according to the normal distribution with the correct answer mean and standard deviation of 0.5.

User 5 (Normal distribution with high variance) answers according to the normal distribution with the correct answer mean and standard deviation of 1.5.

Table 1 summarizes the simulation results, including the type of participation of each user. If the rate of valid answers of a user is more than 50% (meaning that they have answered at least eight questions validly), the user is considered non-Malicious. Otherwise, if the rate of outlier answers of the user is more than 50%, the user is considered malicious. According to the table, the user who answers questions randomly, the user who answers questions according to a specific pattern, and the user who answers questions according to the normal distribution with the correct values mean and standard deviation of 1.5, have a lower number of valid answers than the threshold and are correctly detected as malicious. On the other hand, the accurate user, as well as the user who answers according to the normal distribution with the correct values mean and standard deviation of 0.5, are considered non-Malicious. Eventually, malicious users will be removed from the system, and their responses will not be included in the aggregation process.

TABLE I. SIMULATION RESULTS

| *User* | *User type* | *Number of valid answers* | *Total number of questions* | *participation type* |
|---|---|---|---|---|
| User1 | Random | 5 | 15 | Malicious User |
| User2 | Pattern | 6 | 15 | Malicious User |
| User3 | Accurate | 15 | 15 | non-Malicious User |
| User4 | Normal distribution with low variance | 10 | 15 | non-Malicious User |
| User5 | Normal distribution with high variance | 6 | 15 | Malicious User |

## CONCLUSION

In this paper, at first, the information needs assessment of the data collection system from the flooded area has been performed, and the needs have been categorized. Then the mobile crowdsourcing system provides a questionnaire including multiple-choice and descriptive questions to the users available in the area. The data sent by users are then processed using the proposed malicious user detection algorithm to detect outlier data and malicious users, and subsequently, remove the malicious users from the system. The simulation results via MATLAB have revealed that the proposed scheme reasonably detects outlier data and malicious user data. As a result, organizations that need this information will be more confident in using the information they receive.